\documentclass[pra,twocolumn,english,floatfix,superscriptaddress,nofootinbib]{revtex4-2}

\usepackage[english]{babel}
\usepackage{textcomp,xcolor,natbib}
\usepackage{dcolumn}
\usepackage{graphicx}
\graphicspath{{Figures/}}

\usepackage{amsmath, bbm}
\usepackage{latexsym}
\usepackage{amsfonts}   
\usepackage{amssymb}
\usepackage{array}      
\usepackage{epsfig}
\usepackage{txfonts}
\usepackage{xcolor,braket}
\usepackage[colorlinks=true,linkcolor=blue,urlcolor=blue,citecolor=blue]{hyperref}
\usepackage[normalem]{ulem}
\usepackage{soul}
\usepackage{dsfont}

\usepackage{xfrac}  
\usepackage{relsize}

\usepackage{amsthm}
\usepackage{mathrsfs, mathtools, amsmath}

\usepackage[normalem]{ulem}
\usepackage{cancel}
\usepackage{enumitem}

\usepackage{xspace}

\newcommand{\ketbra}[1]{{\ket{#1}\bra{#1}}}
\newcommand{\id}{{\mathbbm{1}}}
\newcommand{\abs}[1]{\left\lvert #1 \right\rvert}
\newcommand{\norm}[1]{\left\lVert #1 \right\rVert}
\newcommand{\tr}{{\operatorname{tr}}}

\newcommand{\shro}{Schr\"odinger\xspace}
\newcommand{\schro}{\shro}
\newcommand{\heis}{Heisenberg\xspace}
\newcommand{\pdet}{p_{\text{det}}}
\newcommand{\pdis}{p_{\text{d}}}
\newcommand{\pcopy}{p_{\text{c}}}
\newcommand{\pno}{p_{\text{n}}}
\newcommand{\psidet}{\psi_{\text{det}}}

\definecolor{lavender}{rgb}{0.75, 0.58, 0.89}

\begin{document}
\title{Stochastic unravelings for Heisenberg picture and trace-non-preserving dynamics}

\author{Federico Settimo}
\email{fesett@utu.fi}
\affiliation{Department of Physics and Astronomy,
University of Turku, FI-20014 Turun yliopisto, Finland}

\author{Kimmo Luoma}
\affiliation{Department of Physics and Astronomy,
University of Turku, FI-20014 Turun yliopisto, Finland}

\author{Dariusz Chru\'sci\'nski}
\affiliation{Institute of Physics, Faculty of Physics, Astronomy and Informatics,
Nicolaus Copernicus University, Grudziadzka 5/7, 87-100 Toru\'{n},
Poland}

\author{Bassano Vacchini}
\affiliation{Dipartimento di Fisica ``Aldo Pontremoli'', Universit{\`a} degli Studi di Milano, Via Celoria 16, I-20133 Milan, Italy}
\affiliation{Istituto Nazionale di Fisica Nucleare, Sezione di Milano, Via Celoria 16, I-20133 Milan, Italy}

\author{Andrea Smirne}
\affiliation{Dipartimento di Fisica ``Aldo Pontremoli'', Universit{\`a} degli Studi di Milano, Via Celoria 16, I-20133 Milan, Italy}
\affiliation{Istituto Nazionale di Fisica Nucleare, Sezione di Milano, Via Celoria 16, I-20133 Milan, Italy}

\author{Jyrki Piilo}
\email{jyrki.piilo@utu.fi}
\affiliation{Department of Physics and Astronomy,
University of Turku, FI-20014 Turun yliopisto, Finland}

\begin{abstract}
    Stochastic unravelings allow to efficiently simulate open system dynamics, yet their application has traditionally been restricted to master equations {in the \schro picture, which} preserve both Hermiticity and trace.
    In this work, we introduce a general framework that extends piecewise-deterministic unravelings to {the \heis picture and} arbitrary trace-non-preserving master equations, requiring only positivity and Hermiticity of the dynamics.
    Our approach includes, as special cases, unravelings of arbitrary dynamics in the Heisenberg picture, evolutions interpolating between fully Lindblad and non-Hermitian Hamiltonian generators, and equations employed in the derivation of full counting statistics, for which we show it can be used to obtain the moments of the associated probability distribution.
    The framework is suitable for both trace-decreasing and trace-increasing processes through stochastic disappearance and replication of the stochastic realizations, and it is compatible with different unraveling schemes and with reverse jumps in the non-Markovian regime.
\end{abstract}

\maketitle


\section{Introduction}
\label{sec:intro}

Stochastic unravelings are a powerful tool to simulate open quantum system dynamics, in which the time evolution of the system state is reconstructed by averaging over different realizations of a stochastic process on the set of pure states.
Unravelings can be divided into two major families: {they can either consist of {piecewise-deterministic} processes interrupted by random jumps \cite{Dum1992, Dalibard-MCWF, Mlmer1993MonteOptics, Plenio-jumps-review, Breuer1998, Breuer1999, Breuer2004, Piilo-NMQJ-PRL, Piilo-NMQJ-PRA, Smirne-W, Chruscinski-Quantum-RO, Donvil2022, Settimo2024, Settimo2025-SSE} or they can be diffusive \cite{Gisin1992, Diosi-NMQSD, Percival1999, Budini2015, Caiaffa-W-diffusive}.
Several different schemes of stochastic unravelings have been proposed in the literature, including methods that allow to describe non-Markovian dynamics \cite{Diosi-NMQSD, Piilo-NMQJ-PRL, Piilo-NMQJ-PRA, Settimo2024}.}

{So far, stochastic unravelings have been mostly used to simulate open system dynamics represented in the \schro picture, due to the fact that the relevant master equations (MEs) are trace-preserving (TP) \cite{Breuer-Petruccione, Vacchini2024} and therefore naturally allow for an interpretation as average over pure states.
However, open system can be also described in the \heis picture, for which the dynamics is described by a unital but not necessarily TP map \cite{Alicki2007, schro-heis}.
Methods to extend unravelings to the \heis picture have been proposed only in the special case of Gaussian systems \cite{Li2025} or by mimicking unravelings for computing two-time correlations \cite{Breuer1998HeisenbergUnraveling}.
In this work, we provide a general method to extend them to general MEs in the \heis picture.}

{Our novel method, does not apply only to \heis picture MEs, but to arbitrary trace-non-preserving (TNP) MEs preserving positivity and Hermiticity.
Dynamics of this form are widely used in the literature.
For instance, they are employed} in the study of {dynamics generated by non-Hermitian Hamiltonians} {featuring} exceptional points, in which {trace-non-preserving} MEs allow to interpolate between purely non-Hermitian and Lindblad dynamics \cite{Minganti2020, Nori2025}.
TNP MEs also appear naturally in the context of counting fields and full counting statistics \cite{Bagrets2003, Flindt2005, Garrahan2010, Garrahan2011}, allowing one to derive the moments and cumulants of the probability distribution of the investigated quantity.
{Our new unraveling method will therefore extend the domain of applicability of stochastic methods {including in particular} dynamics in the \heis picture.}

{The rest of the paper proceeds as follows.
In Sec.~\ref{sec:MEs} we {briefly review} open system dynamics, both in the \schro and in the \heis picture, {considering general TNP MEs and some unraveling techniques widely used to solve TP \schro picture MEs.}
In Sec.~\ref{sec:unravelings-Heis}, we present our method for extending unravelings to \heis picture and arbitrary TNP MEs.
In Sec.~\ref{sec:examples}, we provide some implementations and examples of our method by applying them to physically relevant TNP MEs.
Lastly, in Sec.~\ref{sec:conclusions}, we present the conclusions of our work.}

\section{Open system dynamics}
\label{sec:MEs}
In this Section, we provide a brief overview of open quantum system dynamics, both in the \schro and in the \heis picture.
We {further mention} physically relevant scenarios in which TNP MEs are used, with particular emphasis on tilted Lindbadians and full counting statistics.
We {finally} provide an overview of some commonly used unraveling schemes for \schro picture MEs.

\subsection{\schro and \heis picture}
\label{subsec:Schro-Heis}
Under the assumption that the system and the environment are initially uncorrelated, the reduced dynamics of the system in the \schro picture is described by a completely positive (CP) trace preserving (TP) map $\Lambda_t$ such that \cite{Breuer-Petruccione,Vacchini2024}
\begin{equation}
    \label{eq:evol_Schro}
    \rho(t) = \Lambda_t[\rho].
\end{equation}
Alternatively, the dynamics can be described also in the \heis picture, in which operators evolve in time instead of states, i.e. \cite{Alicki2007}
\begin{equation}
    \label{eq:evol_Heis}
    X(t) = \Lambda^*_t[X]
\end{equation}
 and is such that for any state $\rho$ and operator $X$ it holds
\begin{equation}
    \tr\big[X\,\Lambda_t[\rho]\big] = \tr\big[\Lambda^*_t[X]\,\rho].
\end{equation}
The dynamical map $\Lambda^*_t$ is the adjoint of $\Lambda_t$ and is a CP unital map, i.e $\Lambda_t^*(\mathbb{I})=\mathbb{I}$, but is not TP unless also $\Lambda_t$ is unital.

The \schro picture dynamics of Eq.~\eqref{eq:evol_Schro} is the solution of a so-called time-convolutionless or time-local ME \cite{Breuer2012a,Chruscinski2014a}
\begin{equation}
    \label{eq:ME_Schro}
    \frac{d\rho}{dt} = \mathcal L_t^S[\rho] = -i[H,\rho] + \sum_{j}\gamma_j\,L_j\rho L_j^\dagger - \frac12\{\Gamma_S,\,\rho\},
\end{equation}
with the same structure as the Lindblad ME \cite{Gorini1976, Lindblad1976}, but where $\Gamma_S = \sum_j \gamma_j L_j^\dagger L_j$, and all operators $H$, $L_j$ and rates $\gamma_j$ can depend explicitly on time. Note, that the dynamical map in the Schr\"odinger picture $\Lambda_t$ also satisfies  

\begin{equation}
    \frac{d}{dt}\Lambda_t = \mathcal L_t^S \circ \Lambda_t .
\end{equation}
Passing to the Heisenberg picture one obviously gets

\begin{equation}   \label{LS*}
    \frac{d}{dt}\Lambda^*_t =  \Lambda^*_t  \circ \mathcal L_t^{S*} .
\end{equation}
Translating this equation for the master equation for  $ X(t) = \Lambda^*_t[X]$ one finds

\begin{equation}
    \frac{d}{dt} X(t) = \Lambda^*_t  \circ \mathcal L_t^{S*}[X] = \mathcal L_t^H[X(t)] ,
\end{equation}
where we introduced the Heisenberg generator

\begin{equation}
    {\mathcal L_t^H} := \Lambda_t^*\circ \mathcal L_t^{S*}\circ{\left(\Lambda_t^*\right)}^{-1}.
\end{equation}
We stress that in general  the Heisenberg generator $ {\mathcal L_t^H}$ is not the same as the adjoint of the Schr\"odinger one  $ {\mathcal L_t^S}$, and it can always be represented as follows \cite{schro-heis}
\begin{equation}
    \label{eq:ME_Heis}
     \mathcal L_t^H[X] = i[\tilde H,X]+\sum_j\xi_j\,R_j\,X\,R_j^\dagger - \frac12\{\Gamma_H,X\},
\end{equation}
where $\Gamma_H = \sum_j \xi_j R_j R_j^\dagger$, with all rates and operators that can depend on time.
Note, however, that in general the Hamiltonians $H,\tilde{H}$, the jump operators $L_j, R_j$ and the rates $\gamma_j,\xi_j$ are different. Only in the special case of commutative dynamical maps, i.e. when $\Lambda_t \circ \Lambda_s = \Lambda_s \circ \Lambda_t$ for any pair of $t$ and $s$ one has $\mathcal{L}^H = \mathcal{L}^{S*}$ which implies $\tilde{H}=H$, $R_j = L_j^\dagger$, and $\xi_j = \gamma_j$.

A key concept for open system dynamics is that of divisibility \cite{Chruscinski2022a}:
the dynamics is said to be \schro (C)P divisible if the map 
\begin{equation}
    \Lambda_{t,s}^S\coloneqq \Lambda_t\circ\Lambda_s^{-1}
\end{equation}
describing the time evolution from time $s$ to time $t>s$ in the \schro picture is (completely) positive.
Similarly, it is \heis (C)P divisible if the map
\begin{equation}
    \Lambda_{t,s}^H\coloneqq \Lambda_t^*\circ{(\Lambda_s^*)}^{-1} = \Lambda_t^*\circ {\left(\Lambda_{t,s}^S\right)^*}\circ{\left(\Lambda_t^*\right)}^{-1}.
\end{equation}
is (completely) positive.
Divisibility has been widely studied in open systems and connected to memory effects and quantum non-Markovianity \cite{BLP2009, BLP-PRA, RHP, BLPV-colloquim,Megier2020a,Megier2020b} and is not equivalent in the two pictures \cite{schro-heis}.
From the point of view of the MEs, CP divisibility corresponds to positivity of the rates $\gamma_\alpha\ge0$ or $\xi_\alpha\ge0$, while P divisibility is equivalent to the weaker condition \cite{Kossakowski-necessary}
\begin{equation}
    \label{eq:P-div_condition}
    \sum_\alpha\gamma_\alpha\abs{\braket{\varphi_\mu\vert L_\alpha\vert\varphi_{\mu^\prime}}}^2\ge0
\end{equation}
for all orthonormal bases $\{\varphi_\mu\}_\mu$ and for all $\mu\ne\mu^\prime$, and similarly for \heis P divisibility.

\subsection{Trace-non-preserving MEs}
\label{subsec:TNP}
In many physically relevant scenarios, one is interested in MEs which preserve positivity but in general not the trace.
A generic ME with these properties can be written in a similar form as Eq.~\eqref{eq:ME_Schro} but with an arbitrary self-adjoint operator in the anticommutator, i.e.
\begin{equation}
    \label{eq:TNP_ME}
    \frac{d\rho}{dt} = \mathcal L[\rho] = -i[H,\rho] + \sum_{j}\gamma_j\,L_j\rho L_j^\dagger - \frac12\{\Gamma,\,\rho\},
\end{equation}
for any $\Gamma=\Gamma^\dagger\ge0$.
Such ME preserves the trace iff $\Gamma =\sum_j\gamma_j L_j^\dagger L_j=\Gamma_S$, corresponding to the \schro picture ME \eqref{eq:ME_Schro}.
In the general case, the time evolution of the trace reads
\begin{equation}
    \frac d{dt}\tr[\rho] = \tr\mathcal L[\rho]= \tr\left[\left(\Gamma_{S}-\Gamma\right)\rho\right].
\end{equation}

MEs of this form are widely used in many physically relevant scenarios.
They encompass the case of non-Hermitian {Hamiltonians, which} have been widely studied both theoretically \cite{Minganti2019, Pick2019, Kumar2025, Heiss2012, El-Ganainy2018, Miri2019, Heiss2004, Bender2007} and experimentally \cite{Ozdemir2019, Chen2021, Zhang2025}.
TNP MEs allow to interpolate between {non-Hermitian and Lindblad dynamics}, and schemes to experimentally obtain such dynamics have been proposed both in the trace-decreasing \cite{Minganti2020} and in the trace-increasing \cite{Nori2025} case, via suitable postselection schemes.
Unraveling schemes for the trace decreasing case have been proposed by dealing with non-linear MEs \cite{Liu-non-linear-ME}, or with stochastic equations that do not preserve the norm of the state vectors \cite{Gisin1992, Dalibard-MCWF}.
However, such unravelings cannot be directly used for arbitrary TNP MEs.

Other examples in which TNP MEs naturally appear encompass situations in which the loss of particles is taken into account \cite{Leon-Montiel2014, Fassioli2010}, in energy transfer scenarios \cite{Rebentrost2009, Kurt2020} as well as in connection to non-Markovianity in the presence of indefinite causal order \cite{Karpat2024, Maity2024}.
A special class of TNP MEs are generated by the so-called tilted Lindbladians \cite{Perfetto2022}, i.e. TNP MEs of the form
\begin{equation}
    \label{eq:tilted_L}
    \frac{d\rho}{dt} = -i[H,\rho] + \sum_{j}e^{-s_j}\gamma_j\,L_j\,\rho\, L_j^\dagger - \frac12\{\Gamma,\,\rho\},
\end{equation}
with $\Gamma = \sum_j\gamma_j L_j^\dagger L_j$.
MEs of this form are widely used to to derive the full counting statistics \cite{Bagrets2003, Flindt2005, Garrahan2010, Garrahan2011, Brange2019, Portugal2023} of a suitable observable or first-passage times \cite{Menczel2026a}.

\subsection{Stochastic unravelings for trace preserving MEs}
\label{subsec:unravelings}
We now provide a brief overview of some commonly used unraveling techniques, which allow for efficient numerical solution of TP MEs in the \schro picture.
Such methods rely on the fact that the exact solution of the ME \eqref{eq:ME_Schro} can always be written as the average over pure states
\begin{equation}
    \label{eq:avg_ME}
    \rho(t) = \sum_i\frac{N_i(t)}{N}\ketbra{\psi_i(t)},
\end{equation}
where $N_i(t)$ is the number of realizations in the state $\psi_i(t)$ and $N=\sum_iN_i(0)$.
Different stochastic methods correspond to different ways of evolving the stochastic vectors $\ket{\psi_i(t)}$.

\subsubsection{Monte-Carlo Wave Function}
\label{subsubsec:MCWF}

A first and widely used unraveling technique is the so-called Monte-Carlo wave function (MCWF) unraveling \cite{Dum1992, Dalibard-MCWF, Mlmer1993MonteOptics, Plenio-jumps-review}.
Let $\psi$ denote the current state of the stochastic realization; the MCWF method is a PDP in which the jumps are of the form
\begin{equation}
    \ket\psi\mapsto\frac{L_j\ket\psi}{\norm{L_j\ket\psi}}\eqqcolon\ket{\psi_j}
\end{equation}
happening {between time $t$ and $t+dt$} with probability
\begin{equation}
    \label{eq:p_j}
    p_j = \gamma_j\, \norm{L_j\ket\psi}^2\, dt.
\end{equation}
The deterministic evolution is given by
\begin{equation}
    \label{eq:det_evo}
    \ket\psi\mapsto \frac{(\id - i\,K\,dt)\ket\psi}{\norm{(\id - i\,K\,dt)\ket\psi}}\eqqcolon\ket{\psidet},
\end{equation}
happening with probability
\begin{equation}
    \label{eq:p_det}
    \pdet = \norm{(\id - i\,K\,dt)\ket\psi}^2,
\end{equation}
{where we introduced} the non-Hermitian effective Hamiltonian
\begin{equation}
    K = H -\frac i2\Gamma_S.
\end{equation}
Notice that the MCWF method can be applied only when all rates $\gamma_j$ are non-negative at all times, otherwise it would lead to unphysical negative probabilities.
Such condition is equivalent to CP divisibility of the dynamical map.

Nevertheless, it is possible to extend the method to temporarily negative rates via the non-Markovian Quantum Jumps (NMQJ) method \cite{Piilo-NMQJ-PRL, Piilo-NMQJ-PRA}, in which the reverse jump $\ket{\psi_i} = L_j\ket{\psi_{i^\prime}}\mapsto \ket{\psi_{i^\prime}}$ happens with probability
\begin{equation}
    \label{eq:p_j_NMQJ}
    p_j^{\text{rev}} = -\frac{N_{i^\prime}}{N_i}\gamma_j\, \norm{L_j\ket{\psi_{i^\prime}}}^2\, dt.
\end{equation}
Notice, however, that these reverse jumps impose mutual dependence among the different stochastic realizations, thus making the simulations less efficient since one needs to store and evolve all trajectories simultaneously.


\subsubsection{Rate Operator}
\label{subsubsec:RO}
Another unraveling scheme is the so-called Rate Operator (RO) unravelings \cite{Settimo2024, Settimo2025-SSE}, which, unlike the MCWF, can deal with non-(C)P-divisible dynamics without requiring mutual dependence among the stochastic realization, thus improving the numerical efficiency of the unravelings.

In order to derive the corresponding PDP, one needs to introduce the RO \cite{Diosi-orthogonal-jumps, Smirne-W, Chruscinski-Quantum-RO}
\begin{equation}
    \begin{split}
        \mathcal R_\psi &\coloneqq \sum_i\gamma_j L_j\ketbra\psi L_j^\dagger+\frac12\left(\ket{\phi_\psi}\bra\psi + \ket\psi\bra{\phi_\psi}\right)\\
        &= \sum_\alpha \lambda_\alpha^\psi \ketbra{\chi^\psi_\alpha}
    \end{split}
\end{equation}
where $\psi$ is the current state of the realization and $\phi_\psi$ is an arbitrary and possibly unnormalized state vector depending on $\psi$; the last equality is the spectral decomposition of $\mathcal R_\psi$.
The jumps are defined as $\ket\psi\mapsto\ket{\chi_\alpha^\psi}$, where $\chi_\alpha^\psi$ are the eigenstates of $\mathcal R_\psi$, and happen with probability
\begin{equation}
    \label{p_j_RO}
    p_\alpha^R = \lambda_\alpha^\psi\,dt,
\end{equation}
where $\lambda_\alpha^\psi$ is the corresponding eigenvalue.
In \cite{Settimo2024}, it was shown that it is always possible to find a transformation $\phi_\psi$ such that the eigenvalues are always positive (and therefore the jump process is well defined) whenever the dynamics is P divisible and also in some cases in which P divisibility is violated.
The deterministic evolution is given by
\begin{equation}
    \label{eq:det_evo_RO}
    \ket\psi\mapsto \frac{(\id - i\,K_\psi\,dt)\ket\psi}{\norm{(\id - i\,K_\psi\,dt)\ket\psi}}\eqqcolon\ket{\psidet^R},
\end{equation}
happening with probability
\begin{equation}
    \label{eq:p_det_RO}
    \pdet^R = \norm{(\id - i\,K_\psi\,dt)\ket\psi}^2,
\end{equation}
with the non-linear effective Hamiltonian
\begin{equation}
    K_\psi = H -\frac i2\Gamma_L-\frac i2\ket{\phi_\psi}\bra\psi.
\end{equation}
In the case of negative eigenvalues, the RO unraveling technique can also be equipped with reverse jumps in a similar way as the MCWF.

\section{Stochastic unravelings in the \heis picture}
\label{sec:unravelings-Heis}

\begin{figure}
    \centering
    \includegraphics[width=0.9\linewidth]{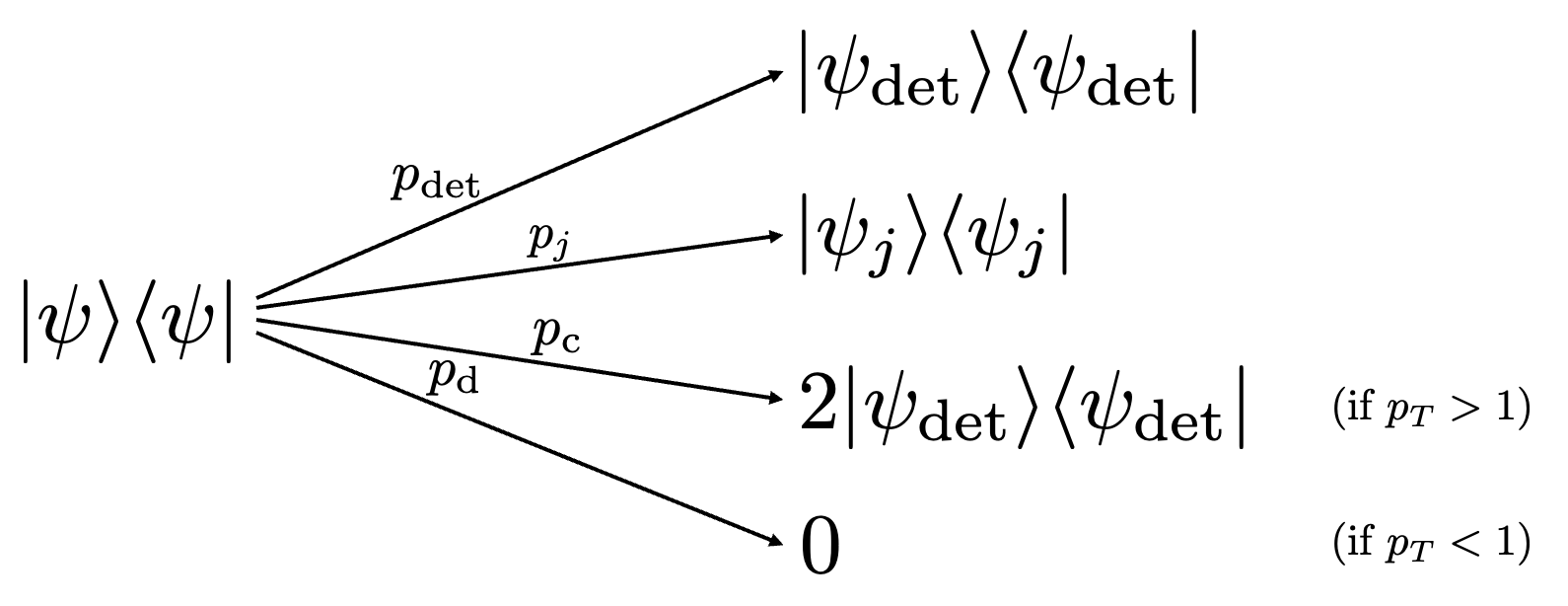}
    \caption{At any given moment of time, there are 4 possibilities for the evolution of the realization, as given in Eqs.~\eqref{eq:prob_all_1}-\eqref{eq:prob_all_2}.
    It can either evolve deterministically, jump, create a copy of itself or vanish.}
    \label{fig:sketch_TNP}
\end{figure}

In this Section, we provide a simple way to extend the stochastic unravelings of Sec.~\ref{subsec:unravelings} from the \schro picture to the \heis picture.
Such method can also be generalized to arbitrary forms of TNP MEs and it coincides with the so-called cloning algorithm when applied to tilted Lindbladians \cite{Giardina2006, Dean2009, Carollo2020}.

In the \schro picture unravelings, trace-preservation via the averaging of Eq.~\eqref{eq:avg_ME} holds since the total number of stochastic realizations is preserved $\sum_iN_i(t)=N$.
Therefore, in order to  generalize the unravelings to TNP MEs such as the \heis picture ME \eqref{eq:ME_Heis} {while keeping the normalization of the state vector, one needs to} allow for the total number of realizations to vary in time.
{The evolution of the trace is then} given by the ratio between the number of realizations at time $t$ and the number at the initial time: $\tr X(t) = \sum_iN_i(t)/N\ne1$.

Another issue of unraveling MEs in the \heis picture is that they act on arbitrary operators which don't need to be positive nor self-adjoint, and therefore cannot be written as an average as in Eq.~\eqref{eq:avg_ME}.
This issue is easily tackled in a similar manner as done in \cite{Settimo2025} for the extension of unravelings to initially correlated system and environment: an arbitrary operator $X$ can first be separated in Hermitian and anti-Hermitian part $X = X_h+iX_a$, $X_{h,a}^\dagger=X_{h,a}$, each of which can be written as the difference between two positive operators $X_{h,a} = X_{h,a}^+-X_{h,a}^-$, where each $X_{h,a}^\pm$ is positive and therefore proportional to a density matrix $X_{h,a}^\pm=\mu_{h,a}^\pm\rho_{h,a}^\pm$.
This way of separating $X$ into the sum of positive operators is preserved by the evolution, since it preserves positivity and Hermiticity.
Therefore, each $\rho_{h,a}^\pm$ is unraveled separately and the solution of the \heis picture ME $X(t)$ is obtained as the weighted sum of the different terms
\begin{equation}
    X(t) = \mu_h^+\rho_h^+(t)-\mu_h^-\rho_h^-(t) + i\mu_a^+\rho_a^+(t) - i\mu_a^-\rho_a^-(t).
\end{equation}

We now show how to generalize unraveling to the \heis picture with a non-constant number of realizations first with the MCWF and then with the RO, with an illustration of our method presented in Fig.~\ref{fig:sketch_TNP}.
However, our method can be generalized in a straightforward way also to different stochastic methods.
Notice that the jumps and deterministic evolution are now defined via the operators $\tilde H$, $R_j$, and $\Gamma_H$ of the \heis picture ME \eqref{eq:ME_Heis}.

\subsection{\heis unravelings with MCWF}
\label{subsec:unravelings-Heis-MCWF}

When considering the MCWF unraveling applied to the operators $\tilde H$, $R^\dagger_j$, $\Gamma_H$ and rates $\xi_j$ of the \heis picture ME \eqref{eq:ME_Heis}, then the jump and deterministic probabilities no longer sum to one{, and therefore do not form a proper probability distributions.
Instead,}
\begin{equation}
    \label{eq:p_T}
    p_T=\pdet +p_J = 1+dt\,\braket{\Gamma_{\text{tp}}-\Gamma_H}_\psi\ne1,
\end{equation}
where $\braket{X}_\psi = \braket{\psi\vert X\vert\psi}$, $p_J \coloneqq \sum_jp_j$ denote the total jump probability, and
\begin{equation}
    \Gamma_{\text{tp}}\coloneqq \sum_\alpha\xi_\alpha\,R_\alpha^\dagger R_\alpha
\end{equation}
is the operator that would make the ME \eqref{eq:ME_Heis} TP.

The {parameter $p_T$, depending on the ME,} can generally be smaller or greater than $1$, with the difference corresponding to the change in trace for the particular realization
\begin{equation}
    p_T-1 = dt\,\frac d{dt}\tr\big[\ketbra\psi\big].
\end{equation}
{Notice that if one particular realization gives $p_T<1$ ($p_T>1$), it does not imply that the same holds also for other trajectories,} nor a decrease (increase) of trace for the average state $\rho$.

Let us first consider the case $p_T<1$.
Since the sum of jump and deterministic probability is smaller than one, we can think of adding an extra process happening with the remaining probability.
Since the trace is decreasing, it is natural for this process to be the disappearance of the stochastic realization $\ket\psi\mapsto 0$ with probability
\begin{equation}
    \pdis = 1-p_T = \braket{\Gamma_H-\Gamma_{\text{tp}}}_\psi\,dt \ge 0.
\end{equation}
Therefore, the realization $\psi$ can either jump, evolve deterministically, or disappear and therefore not counting anymore in the average of Eq.~\eqref{eq:avg_ME}.
The corresponding probabilities are positive and correctly normalized to one.

Indeed, with this extra process, the stochastic unravelings do match the exact solution on average
\begin{equation}
    \begin{split}
        \ketbra\psi \mapsto& \sum_j p_j  \frac{R_j\ketbra\psi R_j^\dagger}{\norm{R_j\ket\psi}^2} \,dt + \pdet{\ketbra\psidet} + \pdis\cdot0\\
        =&\sum_j \xi_j {R_j\ketbra\psi R_j^\dagger}\, dt\\
        &+ (\id - i K\,dt)\ketbra\psi(\id + i K^\dagger\,dt)\\
        =& \ketbra\psi + \mathcal L_t^H[\ketbra\psi]\,dt.
    \end{split}
\end{equation}

For the case $p_T>1$, one cannot {indeed} add an extra event, since $\pdet+p_J$ larger than $1$ and therefore they do not correspond to a proper probability distribution.
{Instead, } we include a new process, corresponding to the creation of a new copy in the same state $\psi$.
Such process happens with probability
\begin{equation}
    \pcopy=p_T-1 = \braket{\Gamma_{\text{tp}}-\Gamma_H}_\psi\,dt \ge 0
\end{equation}
and is statistically independent of the jump or deterministic evolution.
Accordingly, the probability of not creating a copy is $\pno = 1-\pcopy$.
Once one has evaluated whether a copy is created or not, the original realization $\psi$ can evolve either via a jump or deterministically, but the corresponding probabilities must be renormalized in order to sum to 1 as
\begin{equation}
    p_j^\prime = p_j,\qquad \pdet^\prime = 1-p_J.
\end{equation}
Notice that the two copies evolve independently and that the event in which the creation of a copy is followed by a jump happens with probability $O(dt^2)$ and, as it is usually the case for stochastic unravelings, we only compute terms up to $dt${, so one can assume that new copy is created in the state $\psidet$.}
Therefore, the three possible evolutions for $\psi$ are
\begin{enumerate}
    \item no copy and deterministic: $\ket\psi\mapsto\ket\psidet$ with probability $\pno^{\text{det}}=\pno\pdet^\prime = 1-p_J-\braket{\Gamma_{\text{tp}}-\Gamma_H}_\psi\,dt+O(dt^2)$;
    \item no copy and jump: $\ket\psi\mapsto\ket{\psi_j}$ with probability $\pno^j = \pno p_j = p_j + O(dt^2)$;
    \item copy and deterministic: the original copy evolves deterministically and a new copy is created in the ensemble
    \begin{equation}
        {\ketbra\psi\mapsto2\,\ketbra\psidet}
    \end{equation}
    with probability $\pcopy^{\text{det}}=\pcopy\pdet^\prime = \braket{\Gamma_{\text{tp}}-\Gamma_H}_\psi\,dt+O(dt^2)$.
\end{enumerate}
Indeed, also in this case the average evolution matches the solution of the ME \eqref{eq:ME_Heis}:
\begin{equation}
    \begin{split}
        \ketbra\psi\mapsto & \sum_j p_j  \frac{R_j\ketbra\psi R_j^\dagger}{\norm{R_j\ket\psi}^2} \,dt+ \pcopy^{\text{det}}\ketbra\psi\\
        &+ (\pno^{\text{det}}+\pcopy^{\text{det}}){\ketbra\psidet}.
    \end{split} 
\end{equation}
By using $\pno^{\text{det}}+\pcopy^{\text{det}} = \pdet^\prime = 1-\braket{\Gamma_{\text{tp}}}_\psi\, dt$ and expanding the normalization factor of the deterministic term $ 1/\norm{(\id - i K\,dt)\ket\psi}^2$ to the first order in $dt$, it is straightforward to show that
\begin{equation}
    \ketbra\psi\mapsto\ketbra\psi+\mathcal L_t^H[\ketbra\psi]\,dt
\end{equation}
also in the case in which the trace increases $p_T>1$.
{Therefore, by taking the continuous time limit $dt\to 0$, it follows that the evolution of an arbitrary operator $X(t)$ does indeed correspond to the \heis picture ME \eqref{eq:ME_Heis}.}

The correct behavior of the trace on average follows from the linearity of the trace and the fact that each realization evolves on average as prescribed by the ME.

It is possible to merge the two cases $p_T<1$ and $p_T>1$ as
\begin{gather}
    \label{eq:prob_all_1}
    \pdet=1-p_J-\abs{\braket{\Gamma_{\text{tp}}-\Gamma_H}_\psi}\,dt,\\
    \pdis=\max\{0,\braket{\Gamma_H-\Gamma_{\text{tp}}}_\psi\,dt\},\\
    \label{eq:prob_all_2}
    \pcopy=\max\{0,\braket{\Gamma_{\text{tp}}-\Gamma_H}_\psi\,dt\},
\end{gather}
and $p_j$ as in Eq.~\eqref{eq:p_j}; $p_j$ and $\pdet$ correspond to no creation nor disappearance of any trajectory, $\pdis$ describes the disappearance, and $\pcopy$ the creation of a copy plus deterministic evolution for the original copy.
Notice that one of $\pdis$ and $\pcopy$ is always zero and
\begin{equation}
    \sum_jp_j+\pdet+\pdis+\pcopy=1.
\end{equation}
A schematic illustration of the different possibilities is described in Fig.~\ref{fig:sketch_TNP}, {while a specific example is worked out in Sec.~\ref{subsubsec:ex-Heis-Markov}.}

{Notice that it is possible to keep under control the numerical requirements by maintaining a fixed number of realizations in a similar way as done in \cite{Carollo2020}.
This is done by first evolving all trajectories from $t$ to $t+dt$ and then randomly sampling $N(0)$ of the $N(t+dt)$ resulting trajectories and discarding the others.
The change in trace is then simply given by keeping track of the variation $N(t+dt)-N(t)$ but without the need to store more than $N(0)$ trajectories.
In particular, for $N(0)\to \infty$, the fact that only the difference needs to be stored at each time ensures the exact convergence in the asymptotic limit.}

So far, we have assumed that the dynamics is CP-divisible, and therefore all rates $\xi_j$ are positive, and therefore $p_j$ are indeed probabilities.
However, our unraveling method for TNP MEs can be equipped with reverse jumps of Eq.~\eqref{eq:p_j_NMQJ}.
The probabilities for deterministic evolution and for the creation or destruction of a copy are left unchanged and computed using the (possibly negative) $p_j$.
The jump probabilities, instead, are computed using the reverse jumps of Eq.~\eqref{eq:p_j_NMQJ}.
The proof that the unravelings agree with the ME on average is presented in App.\ref{app:proof_NMQJ}.

\subsection{\heis unravelings with RO}
\label{subsec:unravelings-Heis-RO}

The unravelings in the \heis picture can be applied in a simple way also to the RO method of Sec.~\ref{subsubsec:RO}.
To do so, we first notice that although both the jump $p_\alpha^R$ and and deterministic $\pdet^R$ probabilities depend on the transformation $\phi_\psi$, their sum does not
\begin{equation}
    \pdet^R + \sum_\alpha p_\alpha^R = 1+dt\, \braket{\Gamma_{\text{tp}}-\Gamma_H}_\psi = p_T
\end{equation}
and is the same as for the MCWF.
Therefore, it is possible to apply the same ideas also for the RO: the probabilities of creation or destruction of a copy are as in Eqs.~\eqref{eq:prob_all_1}-\eqref{eq:prob_all_2}, but using $p_J^R = \sum_\alpha p^R_\alpha$ instead of $p_J$ and $p^R_\alpha$ as jump probabilities.
Notice that, since $p_T$ is independent of $\phi_\psi$, then also $\pno^R$, $\pcopy^R$ do not depend on it.

For $p_T<1$, the average evolution reads
\begin{equation}
        \ketbra\psi \mapsto \sum_\alpha p_\alpha^\psi\,\ketbra{\chi_\alpha^\psi} + \pdet^R{\ketbra{\psidet^R}} + \pdis^R\cdot0,
\end{equation}
while for $p_T>1$ it reads
\begin{equation}
    \begin{split}
        \ketbra\psi\mapsto& \sum_\alpha p_\alpha^\psi\,\ketbra{\chi_\alpha^\psi} 
        + \pcopy^{\text{det},R}\ketbra\psi\\
        &+ (\pno^{\text{det},R}+\pcopy^{\text{det},R}){\ketbra{\psidet^R}}.
    \end{split} 
\end{equation}
By direct calculation, it is easy to show that in both cases
\begin{equation}
    \ketbra\psi\mapsto\ketbra\psi+\mathcal L_t^H[\ketbra\psi]\,dt
\end{equation}
and indeed the \heis picture RO unravelings match the ME \eqref{eq:ME_Heis}. {This situation is considered in Sec.~\ref{subsubsec:ex-Heis-non-Markov}.}

If the RO has negative eigenvalues, then it is possible to equip the unravelings with the reverse jumps of Eq.~\eqref{eq:p_j_NMQJ} in a similar way as for the MCQF, see App.~\ref{app:proof_NMQJ} for further details.

\subsection{Generic TNP MEs}
\label{subsec:unravelings-TNP-generic}

Our method for extending stochastic unravelings to the \heis picture can be applied in a straightforward way to arbitrary TNP MEs \eqref{eq:TNP_ME}.
This is done by substituting $\Gamma_H$ with $\Gamma$ of Eq.~\eqref{eq:TNP_ME} in Eqs.~\eqref{eq:prob_all_1}-\eqref{eq:prob_all_2}.
The jump probabilities can be then computed with either the MCWF or the RO formalism.
This allows to extend the powerful simulation method that are stochastic unravelings not only beyond the \schro picture, but also to arbitrary TNP MEs. {An example in this direction is given in Sec.~\ref{subsec:photon-counting}.}

When considering MEs corresponding to the statistics of time-integrated observables, our method with the MCWF coincides with the cloning method \cite{Giardina2006, Dean2009, Carollo2020}.
Such method {was devised} for the characterization of trajectory-dependent observables of the form \cite{Carollo2020}
\begin{equation}
    \mathcal O[\omega_t] = \sum_{k=0}^{n-1}\,\alpha\big(\ket{\psi_k},\ket{\psi_{k+1}}\big),
\end{equation}
where $\omega_t = \big\{\ket{\psi_0},\ket{\psi_1},\ldots,\ket{\psi_n}\big\}$ is a (discretized) stochastic trajectory and $\alpha$ is an observable-dependent function.
For the special case of $\alpha$ being non-zero if and only if $\ket{\psi_{k+1}} = L_j\ket{\psi_{k}}$, the resulting dynamics is the tilted Lindbladian of Eq.~\eqref{eq:tilted_L} and the resulting observable is the number of jumps for the trajectory $\omega_t$.

\section{Implementations and examples of TNP unravelings}
\label{sec:examples}

In this Section, we apply our method of stochastic unravelings in the \heis picture described in Sec.~\ref{sec:unravelings-Heis} to two examples, one CP-divisible and one non-CP-divisible.
In Sec.~\ref{subsec:photon-counting}, we apply a variation of the same method to tilted Lindbladians applied to photon counting statistics.

\subsection{\heis picture ME}
\label{subsec:ex-Heis}

In order to exemplify the \heis picture stochastic unravelings we consider two examples.
First, we consider a dynamics which is CP-divisible in the \heis picture but not in the \schro picture, for which the MCWF method can be applied.
Second, we consider a dynamics non-divisible in both pictures, for which one needs to either use the RO or reverse jumps.

\subsubsection{\heis CP-divisible}
\label{subsubsec:ex-Heis-Markov}

As a first example, we consider a \heis picture ME of the form
\begin{equation}
    \label{eq:Heis_ex_ME}
    \begin{split}
        \frac {dX}{dt}=& i\,\omega[\sigma_x,X]+\xi_-\left(\sigma_+X\sigma_- - \frac12\{\sigma_+\sigma_-,X\}\right)\\
        &+ \xi_+\left(\sigma_-X\sigma_+ - \frac12\{\sigma_-\sigma_+,X\}\right),
    \end{split}
\end{equation}
where $\sigma_+ = \ket1\bra0=\sigma_-^\dagger$, $\omega$ and $\xi_{\pm,z}$ are time-dependent functions.
We consider the strongly driven regime $\abs{\omega}\gg\xi_\pm>0$.
{For various choices of time dependent rates, this} corresponds to a dynamics that is CP divisible in the \heis picture ($\xi_\pm\ge0$) but P indivisible in the \schro picture, as shown in App.~\ref{app:no_div_S}.
In particular, {in the strongly-driven regime, the} dynamics is not \schro P nor CP-divisible since the initial time, thus making the reverse jumps fail.
Because of divisibility, the unravelings in the \heis picture can be performed using the MCWF and do not require reverse jumps, while if one unraveled the same dynamics in the \shro picture they would be necessary.
This makes the simulations noticeably more efficient in the \heis picture.

\begin{figure}
    \centering
    \includegraphics[width=\linewidth]{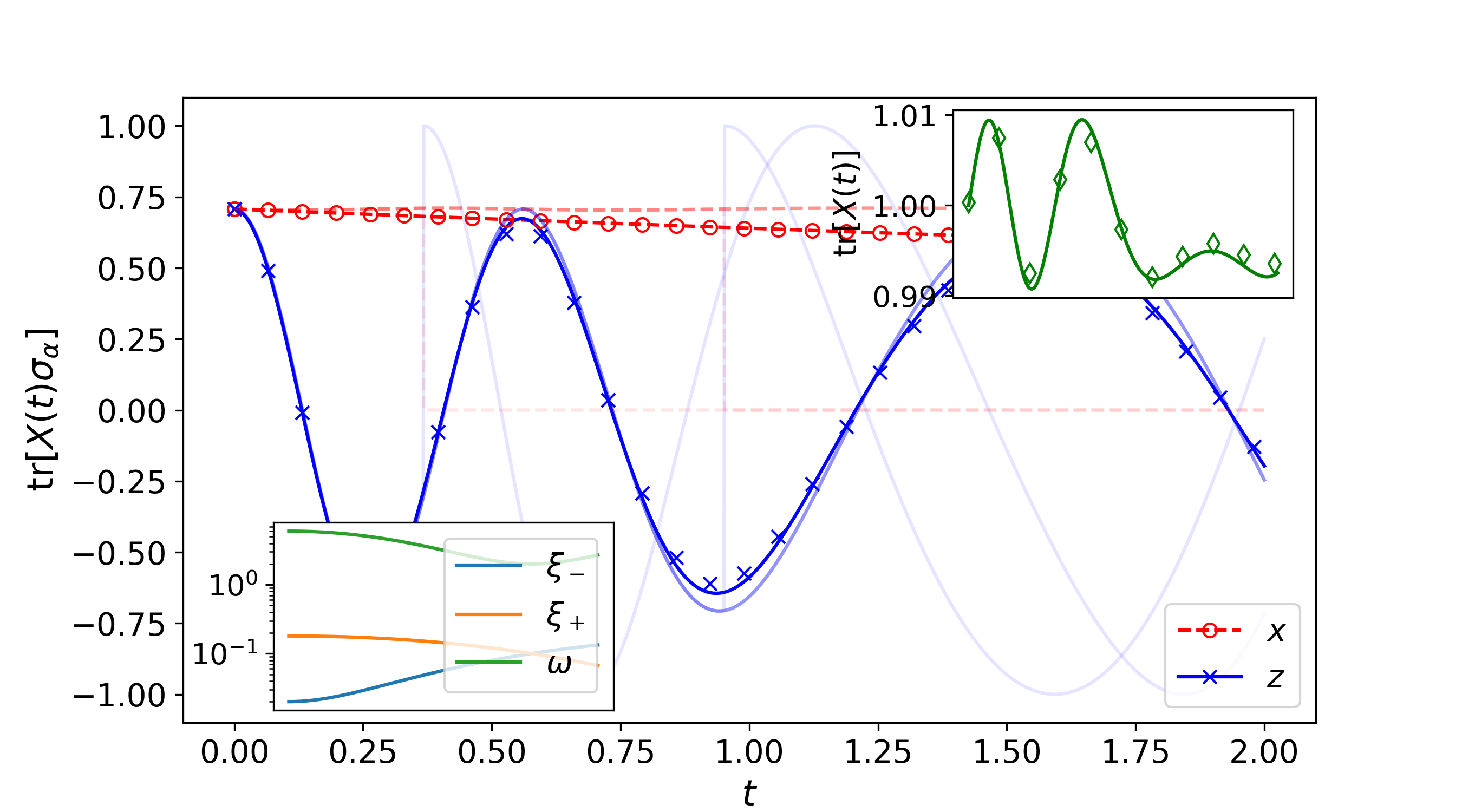}
    \caption{Dynamics of the \heis picture ME \eqref{eq:Heis_ex_ME}, $x$ and $z$ components of the Bloch vector.
    The MCWF unravelings match the exact solution (dark lines); in lighter shade $7$ stochastic trajectories are also shown.
    Lower left inset: rates $\xi_\pm$ and $\epsilon$ (logarithmic scale).
    Upper right inset: dynamics of the trace $\tr[X(t)]$, obtained as the ratio $\sum_iN_i(t)/N$.
    The timestep used in the simulations is $dt = 10^{-3}$; $N=2\cdot10^4$ stochastic realizations are present at $t=0$, the number at time $t$ follows the same dynamics as $\tr[X(t)]$ (upper right inset).
    Initial value $X(0) = \ketbra{\psi_0}$, with $\ket{\psi_0} = \cos(\pi/8)\ket++\sin(\pi/8)\ket-$.}
    \label{fig:Heisenberg}
\end{figure}

In Fig.~\ref{fig:Heisenberg}, we show the agreement between the MCWF unravelings and the exact solution of the \heis picture ME.
In particular, in the upper right panel, we show that the trace of the observable $\tr X(t)$ oscillates between values greater and smaller than its initial value $\tr X(0) = 1$.
This shows that our method not only gives the correct expectation values for any measurement of the form $\tr[\rho\,X(t)]$, but it also correctly captures the time evolution of the trace via the non-constant number of stochastic realizations.
The code used for the simulation is available at \cite{github}.

\subsubsection{\heis non-CP-divisible}
\label{subsubsec:ex-Heis-non-Markov}

As a second example, we consider a \heis picture ME of the form 
\begin{equation}
    \label{eq:Heis_ph_cov}
    \begin{split}
        \frac {dX}{dt}&= \sum_{\alpha=\pm,z}\xi_\alpha\left(\sigma_\alpha X\sigma_\alpha^\dagger - \frac12\{\sigma_\alpha\sigma_\alpha^\dagger,X\}\right),
    \end{split}
\end{equation}
where $\xi_{\pm,z}$ are time-dependent functions.
Such ME describes the phase covariant dynamics \cite{Haase-fundamental, Smirne-ultimate, Vacchini2010-notes} in the \heis picture \cite{schro-heis}.
The resulting dynamics is \heis CP-divisible if all rates $\xi_j$ are positive, while P-divisibility corresponds to \cite{Filippov-ph-cov, Teittinen2021}
\begin{equation}
    \label{eq:ph_cov_P-div}
    \xi_\pm\ge0,\quad\text{and}\quad\xi_z\ge-\frac12\sqrt{\xi_+\xi_-}.
\end{equation}

\begin{figure}
    \centering
    \includegraphics[width=\linewidth]{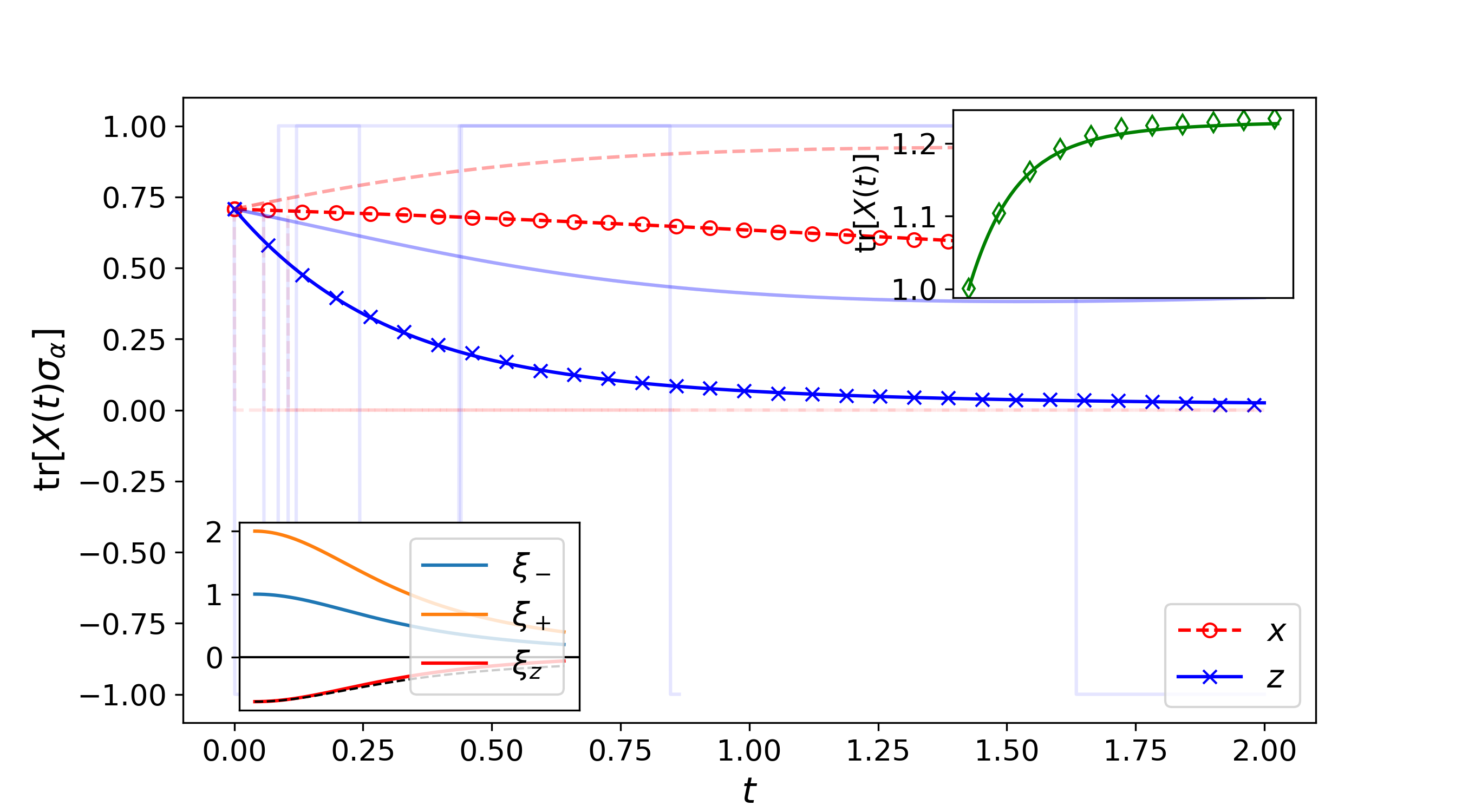}
    \caption{Dynamics of the non-CP-divisible \heis picture ME \eqref{eq:Heis_ph_cov}, $x$ and $z$ components of the Bloch vector.
    The RO unravelings match the exact solution (dark lines), with no reverse jumps required; in lighter shade $7$ stochastic trajectories are also shown.
    Lower left inset: rates $\xi_\alpha$; the dashed line is the condition for P-divisibility for $\gamma_z$ \eqref{eq:ph_cov_P-div}.
    Upper right inset: dynamics of the trace $\tr[X(t)]$.
    The other parameters and initial value are the same as in Fig.~\ref{fig:Heisenberg}.}
    \label{fig:Heisenberg_nM}
\end{figure}

We choose the rates in such a way that CP-divisibility is violated at all times, while P-divisibility is preserved, in a similar manner to the eternally-non-Markovian dynamics \cite{Megier2017, Hall2014}.
The violation of divisibility arises from the negativity of $\xi_z$, while $\xi_\pm$ are positive at all times, see Fig.~\ref{fig:Heisenberg_nM}.
Because of the form of the ME and $\xi_z<0$, the dynamics is non-CP-divisible also in the \schro picture \cite{schro-heis}.

Using the RO to perform the \heis picture unravelings, it is possible to obtain the evolution of any operator $X(t)$ without the need of reverse jumps despite $\xi_z<0$ at all times.
Furthermore, it is possible to perform the unravelings with a small effective ensemble: only the eigenstates $\ket0$, $\ket1$ of $\sigma_z$ and the state conditioned to no jumps happening are needed to describe the reduced dynamics.
These facts are possible thanks to the flexibility of the RO with a trajectory-dependent transformation $\phi_\psi$ and they drastically improve the efficiency of the simulation technique.

The resulting unravelings are presented in Fig.~\ref{fig:Heisenberg_nM} and, unlike the previous example, here the trace $\tr X(t)$ is monotonically increasing in time.

\subsection{Photon counting}
\label{subsec:photon-counting}

\begin{figure}
    \centering
    \includegraphics[width=\linewidth]{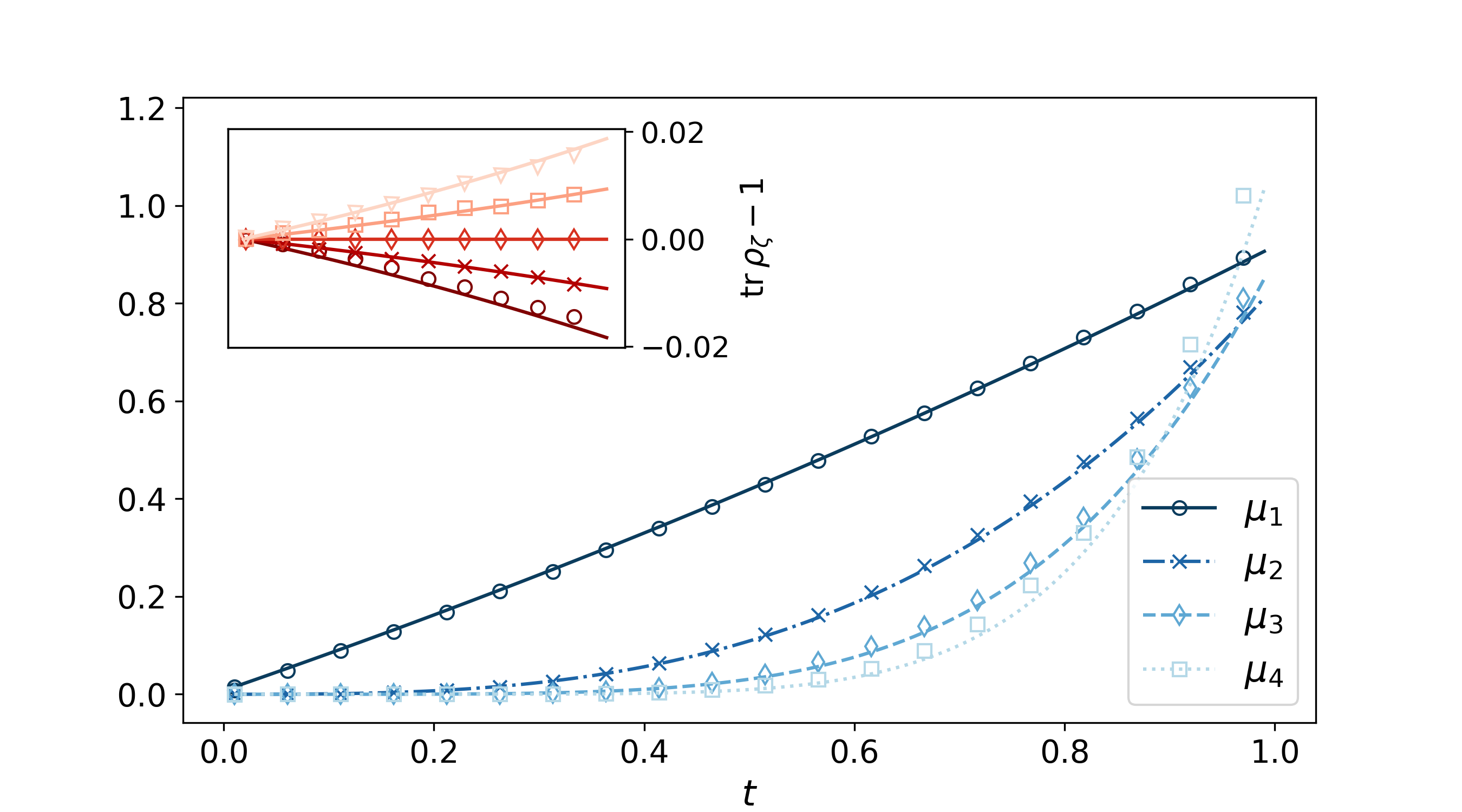}
    \caption{Moments of $P(n,t)$.
    The exact solution is obtained by solving Eq.~\eqref{eq:ME_photon_counting} and taking the derivative of the trace; the unravelings are performed on Eq.~\eqref{eq:ME_tau_k}.
    Inset: $\tr\rho_\zeta - 1$, for $\zeta=-0.02$ (bottom) to $\zeta=0.02$ (top).
    Parameters: $\gamma = \Omega = 1$, $\bar n = 0.5$, $\phi = 0.2$, $dt = 10^{-2}$, initial number of trajectories: $N = 10^4$, initial state: $\ket{\psi_0} = (\ket0+\ket1)/\sqrt{2}$.
    {Notice that the final time is chosen to be comparable to the typical timescale $1/\gamma$ of the dynamics.}}
    \label{fig:photon-counting}
\end{figure}

As a last example, we apply our method to a special case of tilted Lindbladian \eqref{eq:tilted_L}, in which the TNP ME reads \cite{Portugal2023}
\begin{equation}
    \label{eq:ME_photon_counting}
        \frac d{dt}\rho_\zeta = \mathcal L[\rho_\zeta] + \zeta\,\mathcal J[\rho_\zeta],
\end{equation}
where $\mathcal{L}$ is a TP Lindbladian
\begin{equation}
    \mathcal L[\rho] = -i[H,\rho]+\gamma(\bar n+1)a\rho a^\dagger + \gamma\bar n a^\dagger\rho a - \frac12\{\Gamma_{S},\rho\},
\end{equation}
with $a^\dagger$, $a$  bosonic creation and annihilation operators, $\zeta$, $\gamma$, $\bar n$ positive real numbers and Hamiltonian
\begin{equation}
    H = \frac\Omega 2\left(ae^{2i\phi} + a^\dagger e^{-2i\phi}\right),
\end{equation}
where $\phi$ and $\Omega$ are real parameters{, while $\mathcal J[\rho] = \gamma(\bar n+1)a\rho a^\dagger$.}
MEs of this form are typically used to describe the photon counting statistics \cite{Brange2019}.
{In particular, the trace of $\rho_\zeta$ allows to compute the factorial moments of the probability distribution $P(n,t)$ of observing $n$ {photons} at time $t$ as
\begin{equation}
    \mu_k = \frac{d^k}{d\zeta^k}\tr\left[\rho_\zeta\right]\Big\vert_{\zeta=0}.
\end{equation}
In principle, one can directly apply the derivative to the estimate of $\tr\rho_\zeta$ obtained by unraveling the TNP ME \eqref{eq:ME_photon_counting} for values of $\zeta$ around 0, but {the obtained results are too noisy}, since the derivative increases the fluctuations due to the {creation} or destruction of extra copies.
An alternative approach is that of defining $\tau_k\coloneqq d^k\rho_\zeta/d\zeta^k\vert_{\zeta=0}$, which {are the solution of the TNP ME} \cite{Perfetto2022, Zoller1987}
\begin{equation}
    \label{eq:ME_tau_k}
    \frac{d}{dt}\tau_k = \mathcal L[\tau_k] + k\, \mathcal J[\tau_{k-1}]
\end{equation}
and the moments can be obtained as $\mu_k=\tr[\tau_k]$.
This system of MEs can be simulated iteratively, with $\tau_0=\rho$ and using the solution of the $k$-th operator $\tau_{k-1}$ in the ME for $\tau_{k}$.
Notice that this does not impact drastically the numerical requirements, since at each time only the trajectories for $\tau_k$ need to be stored{, as well as the average $\tau_{k-1}$}.
Typically only the first few moments are of interest, which are sufficient for example to determine whether the distribution is Poissonian or not \cite{Gerry2004}.
The resulting MEs are inhomogeneous due to the term {$k\,\mathcal J[\tau_{k-1}]$}, which does not depend on $\tau_k$.
By using its spectral decomposition {$k\,\mathcal J[\tau_{k-1}] = \sum_i \eta_i^k \ketbra{\xi_i^k}$}, our method can be extended to deal with it by adding an extra event, corresponding to the creations of copies in the eigenstates {$\ket{\xi_i^k}$} of the inhomogeneous term, with rate equal to the corresponding eigenvalue {$\eta_i^k$}, which are positive since $\mathcal J$ is CP.
Such extra copies and the rate of creation do not depend on the state $\psi$ of the realization.
The initial value for $k\ge1$ is $\tau_k(0)=0$, and thus initially only the inhomogeneous term {is relevant for} the unravelings.
In Fig.~\ref{fig:photon-counting} we present the first four moments obtained by unraveling the TNP ME \eqref{eq:ME_tau_k} with the RO formalism.
This shows that indeed our method can be applied even beyond MEs in the \heis picture.}

\section{Conclusions}
\label{sec:conclusions}

{In this work, we have introduced a general framework to extend stochastic unravelings beyond the commonly used \schro picture.
We first extended them to the \heis picture, in which, by employing the fact that the {property of divisibility} might be different in the {two pictures}, the simulations can become more efficient.}

{We have then showed that our method can be trivially extended to generic} TNP MEs that preserve positivity and Hermiticity{, thus significantly broadening the scope of stochastic unravelings} beyond the standard Lindblad ME in the \shro picture, enabling simulations of a wide variety of physically relevant dynamics.
This includes, for instance, dynamics interpolating between non-Hermitian and Lindblad regimes, and MEs used to extract photon counting statistics.

While we developed the method within the {MCWF and RO} formalism, our method can be easily applied to other piecewise-deterministic schemes, even in the presence of reverse jumps{, thus providing a versatile tool for simulating fully general open quantum systems dynamics.}

\section*{Acknowledgments}
{The authors thank Federico Carollo, Juan P. Garrahan, and Carlos P\'erez Espigares for very useful discussions and comments.}
FS thanks Pedro Portugal for the useful discussion.
FS acknowledges support from Magnus Ehrnroothin S\"a\"ati\"o.
BV and AS acknowledge support from MUR and Next Generation EU via the PRIN 2022 Project “Quantum Reservoir Computing (QuReCo)” (contract n. 2022FEXLYB)  and the NQSTI-Spoke1-BaC project QuSynKrono (contract n. PE00000023-QuSynKrono).
DC was supported by the Polish National Science Center under Project No. 2018/30/A/ST2/00837. 
The authors thank the Toru\'n group and the Aleksander Jab\l o\'nski Foundation for hospitality received.

\appendix

\section{Proof of the reverse jumps}
\label{app:proof_NMQJ}

The jump rates $\xi_j$ of the \heis picture unravelings can be decomposed into a positive and negative part $\xi_j^\pm = \frac12(\abs{\xi_j}\pm\xi_j)$.
The presence of negative rates does not alter the deterministic nor the creation/disappearance of copies, therefore it is sufficient to show that the jump process reproduces on average the jump term of the ME \eqref{eq:ME_Heis}.
The average jump process reads \cite{Breuer2008}
\begin{equation}
    \begin{split}
        \ketbra\psi\mapsto& dt\,\sum_j\xi^+_j R_j\ketbra\psi R_j^\dagger\\
        &-dt\,\sum_j\int d\psi^\prime\,\xi_{j}^-\frac{N_{\psi^\prime}}{N_\psi}\ketbra{\psi}\cdot\\
        \cdot&\delta\left(\ket\psi-\frac{R_j\ket{\psi^\prime}}{\norm{R_j\ket{\psi^\prime}}}\right)\, \norm{R_j\ket{\psi^\prime}}^2.
    \end{split}
\end{equation}
The first term of r.h.s. is the jump term of the ME for positive rates.
In order to show the agreement on average with the ME, one needs to compute the average jump evolution of the unnormalized state
\begin{equation}
    \label{eq:avg_state_NMQJ_proof}
    \rho = \int d\psi\,\frac{N_\psi}N\ketbra\psi,
\end{equation}
where, unlike for the TP case, $N_\psi/N$ is not a probability distribution since it is not normalized {to $1$}.
The average jump evolution then reads
\begin{equation}
    \begin{split}
        \rho\mapsto& \sum_j\xi^+_j R_j\rho R_j^\dagger\\
        &- dt\,\sum_j\int d\psi\int d\psi^\prime\,\xi_{j}^-\frac{N_{\psi^\prime}}{N}\ketbra{\psi}\cdot\\
        &\cdot\delta\left(\ket\psi-\frac{R_j\ket{\psi^\prime}}{\norm{R_j\ket{\psi^\prime}}}\right)\, \norm{R_j\ket{\psi^\prime}}^2.
    \end{split}
\end{equation}
For the second term, the integral in $\psi$ is readily evaluated thanks to the Dirac delta, and therefore the average evolution due only to the jumps is
\begin{equation}
    \begin{split}
        \rho\mapsto \sum_j\xi_j R_j\rho R_j^\dagger.
    \end{split}
\end{equation}
Therefore, since deterministic evolution, creation and destruction of a copy are not modified by the presence of negative rates, it follows that  the unravelings with NMQJ match with the \heis picture ME also when reverse jumps are taken into account.

In a similar manner, it is possible to show that also the \heis picture RO unravelings can be equipped with the NMQJ.
The RO unravelings can be equipped with NMQJ also for TNP MEs.
If $\lambda_j^\psi<0$, then the reverse jump $\ket\psi\mapsto\ket{\psi^\prime}$, where $\ket\psi$ is an eigenstate of $R_{\psi^\prime}$, can happen with probability
\begin{equation}
    p_\alpha^{\text{rev},R} = -\lambda_\alpha^\psi\frac{N_{\psi}}{N_{\psi^\prime}}\,dt.
\end{equation}
The rest of the process is left unchanged: $\pdet^R$, $\pcopy^R$, and $\pdis^R$ are computed using the (possibly negative) rates $\lambda_j^\psi$.
The RO can be written as $\mathcal R_\psi = \mathcal R_\psi^+-\mathcal R_\psi^-$, with
\begin{equation}
    \mathcal R_\psi^\pm = \sum_\alpha\lambda_{\alpha,\psi}^\pm\ketbra{\chi_\alpha^\psi}\ge0,\quad\lambda_{\alpha,\psi}^\pm = \frac12\left(\abs{\lambda_\alpha^\psi}\pm\lambda_\alpha^\psi\right)\ge0.
\end{equation}
Similarly to the MCWF, it is sufficient to show that the jump process reproduces on average the RO, with average jump process reading \cite{Settimo2025-SSE}
\begin{equation}
    \begin{split}
        \ketbra\psi\mapsto &dt\,\sum_\alpha\lambda_{\alpha,\psi}^+\ketbra{\chi_\alpha^\psi}\\
        &-\int d\psi^\prime\,\lambda_{\alpha,\psi^\prime}^-\frac{N_{\psi^\prime}}{N_\psi}\ketbra{\psi}\delta\left(\ket\psi-\ket{\chi_\alpha^{\psi^\prime}}\right).
    \end{split}
\end{equation}
Upon considering the average state as in Eq.~\eqref{eq:avg_state_NMQJ_proof} and using the Dirac delta to compute the integral of the second term, one finds that the average jump evolution reads
\begin{equation}
    \begin{split}
        \rho\mapsto&\int d\psi\,\frac{N_\psi}N \mathcal R_\psi^+\,dt - \sum_\alpha\int d\psi^\prime\,\lambda_{\alpha,\psi^\prime}^-\frac{N_{\psi^\prime}}{N}\ketbra{\chi^{\psi^\prime}_\alpha}\,dt\\
        =&\int d\psi\,\frac{N_\psi}N\left(\mathcal R_\psi^+-\mathcal R_\psi^-\right)\,dt = \int d\psi\,\frac{N_\psi}N\mathcal R_\psi\,dt\\
        =& \sum_j\xi_j R_j\rho R_j^\dagger\, dt + \frac12\int d\psi\,\frac{N_\psi}N\left(\ket{\phi_\psi}\bra\psi+\ket\psi\bra{\phi_\psi}\right)\, dt.
    \end{split}
\end{equation}
From here, it is straightforward to notice that the last term cancels out with the same term with opposite sign from the deterministic evolution, thus canceling all terms depending on $\phi_\psi$ in the average evolution.
Therefore, $\rho\mapsto\rho+\mathcal L^H_t[\rho]\,dt$ and also for the RO the unravelings match with the \heis picture ME also when reverse jumps are taken into account.

\section{Divisibility of the dynamics of Eq.~\eqref{eq:Heis_ex_ME}}
\label{app:no_div_S}

\begin{figure*}
    \centering
    \includegraphics[width=.9\linewidth]{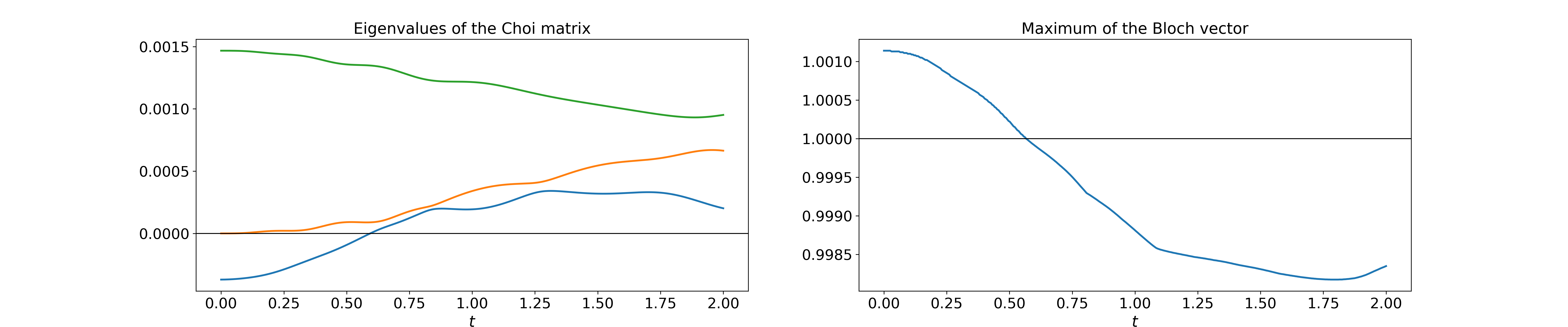}
    \caption{Left panel: the three smallest eigenvalues of the Choi state of Eq.~\eqref{eq:app_Choi}, the fourth eigenvalue is such that $\tr J_t=1$.
    Negativity of one eigenvalue implies violations of CP divisibility.
    Right panel: maximal norm of the Bloch vector under the action of $\Lambda_{t+dt,t}$.
    The fact that such norm is greater than one implies that the dynamics is not P divisible.}
    \label{fig:Choi_Schro}
\end{figure*}

We proceed to show that the solution of the TNP ME \eqref{eq:Heis_ex_ME} gives a dynamical map which is CP-divisible in the \heis picture and P- and CP-indivisible in the \schro picture.

CP-divisibility in the \heis picture follows trivially from the positivity of the rates $\xi_\pm(t)$ of the ME.
Let $\Lambda^*_t$ be the solution of the ME \eqref{eq:Heis_ex_ME}, then the dynamics in the \schro picture is described by its adjoint $\Lambda_t$, which is a CPTP map.
(C)P divisibility in the \schro picture is equivalent to the (complete) positivity of the map $\Lambda_{t+dt,t} = \Lambda_{t+dt}\Lambda_t^{-1}$ for all times $t$.
Complete positivity can be evaluated by checking the positivity of the corresponding Choi state
\begin{equation}
    \label{eq:app_Choi}
    J_t \coloneqq \sum_{i,j=0}^1\Lambda_{t+dt,t}\big[\ket i\bra j\big]\otimes \ket i\bra j.
\end{equation}
Positivity, on the other hand, is equivalent to $\Lambda_{t+dt,t}$, when considered in its Bloch representation, mapping the Bloch sphere into itself, i.e. every vector on the unit sphere must be mapped to vectors with norm smaller than $1$.
In Fig.~\ref{fig:Choi_Schro}, we show that both these conditions are not satisfied.
In particular, they are {violated} since $t=0$, thus making most unraveling methods fail, even in the presence of reverse jumps.

\bibliographystyle{apsrev4-2}
\bibliography{biblio}

\end{document}